\documentclass[twocolumn,amsmath,amssymb,floats,superscriptaddress,nofootinbib,10pt]{revtex4-1}

\pdfoutput=1

\AtBeginDocument{%
    \newwrite\bibnotes
    \def\bibnotesext{Notes.bib}
    \immediate\openout\bibnotes=\jobname\bibnotesext
    \immediate\write\bibnotes{@CONTROL{REVTEX41Control}}
    \immediate\write\bibnotes{@CONTROL{%
    apsrev41Control,author="08",editor="1",pages="1",title="0",year="1"}}
     \if@filesw
     \immediate\write\@auxout{\string\citation{apsrev41Control}}%
    \fi
}%

\usepackage{graphicx}
\usepackage{dcolumn}
\usepackage{bm}
\usepackage{revsymb}
\usepackage[usenames]{color}
\usepackage{color}
\usepackage{physics}
\usepackage{amsmath}
\usepackage[usenames]{color}
\usepackage{amsfonts}
\usepackage{graphicx}
\usepackage{dcolumn}
\usepackage{bm}
\usepackage{revsymb}
\usepackage[usenames]{color}
\usepackage{graphicx}

\usepackage{color}
\usepackage{physics}
\usepackage{amsmath}
\usepackage{amssymb,bbm}
\usepackage{subfig}

\usepackage{xcolor}

\newcommand{\bea}{\begin{eqnarray}}
\newcommand{\eea}{\end{eqnarray}}

\definecolor{nblue}{RGB}{28,130,185}

\definecolor{cgreen}{RGB}{76,153,0}

\definecolor{myorange}{RGB}{245,156,74}

\usepackage{hyperref}
\hypersetup{
  colorlinks=true,
  citecolor=magenta,
  urlcolor=-myorange
}

\usepackage{xcolor}
\definecolor{ogreen} {RGB}{71,191,145}

\usepackage{hyperref}
\hypersetup{
  colorlinks=true,
  citecolor=magenta,
  urlcolor=-myorange
}

\definecolor{oblue} {RGB}{0,0,0}
 
\newcommand{\mj}[1]{\textcolor{oblue}{#1}}

\newcommand{\quotes}[1]{``#1''}

\begin{document}

\title{Slip-induced odd viscous flow past a cylinder}

\author{Ruben Lier}
\email{r.lier@uva.nl}
\affiliation{Institute for Theoretical Physics, University of Amsterdam, 1090 GL Amsterdam, The Netherlands}
\affiliation{Dutch Institute for Emergent Phenomena (DIEP), University of Amsterdam, 1090 GL Amsterdam, The Netherlands}

\begin{abstract}
Odd viscosity is a transport coefficient that can occur when fluids experience breaking of parity and time-reversal symmetry. Previous knowledge indicates that cylinders in incompressible odd viscous fluids, under no-slip boundary conditions, do not exhibit lift force, a phenomenon that poses challenges for the experimental detection of odd viscosity. This study investigates the impact of slip in Stokes flow, employing the odd generalization of the Lorentz reciprocal theorem. Our findings reveal that, at linear order in slip length, lift does not manifest. Subsequently, we explore the scenario involving a thin sheet with momentum decay as well as that of a finite system size, demonstrating that for Stokes flow lift does occur for the second order slip length contribution. We address cylinder flow beyond the Stokes approximation by solving the Oseen equation to obtain a fluid profile that shows an interplay between odd viscosity and inertia, and acquire an explicit expression for Oseen lift at leading order in slip length.
\end{abstract}

\maketitle

\section{Introduction}
\begin{figure*}[t]
    \centering \includegraphics[width=1\linewidth]{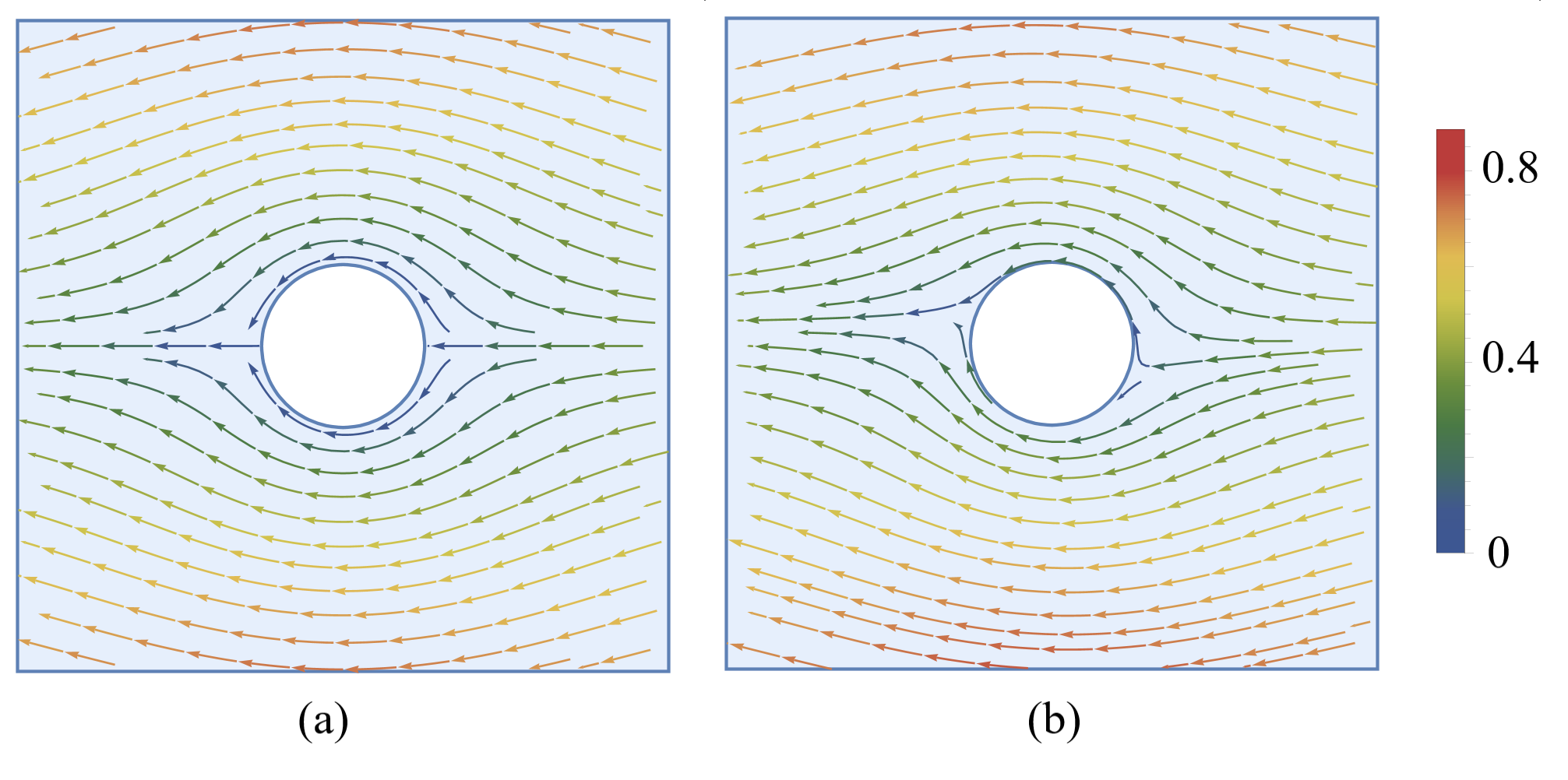}
    \caption{Picture of Oseen flow past a cylinder with odd viscosity for (a) vanishing slip length and (b) a slip length $\hat \lambda = 1$. We took $\gamma_o = 1 $ and $ \hat k = 0.2 $. The arrows and coloring respectively denote the orientation and magnitude of the dimensionless velocity $\hat v_i = v_i/U$.}
    \label{fig:enter-label2981}
\end{figure*}

A question that has concerned fluid physicists for more than a century is how much drag force an object moving in a fluid experiences \cite{veysey2007simple}. The origin of drag force lies with dissipative effects, which can be characterized by phenomenological transport coefficients such as shear viscosity. Over the last decade, there has been a notable focus on a different transport coefficient known as odd viscosity \cite{avron1998odd,avron1995viscosity,levay1995berry}. This transport coefficient is non-dissipative and appears when microscopic two-dimensional chiral effects break parity and time-reversal symmetry, and is important in biological physics~\cite{han2021fluctuating,soni2019odd,markovich2021odd}, electron fluids \cite{pellegrino2017nonlocal,narozhny2019magnetohydrodynamics,berdyugin2019measuring}, topological waves \cite{souslov2019topological,tauber2019bulkinterface,monteiro2023coastal}. Furthermore, odd viscous flow in general has also been a source of many novel fluid mechanical problems~\cite{banerjee2017odd,PhysRevFluids.7.043301,abanov2018odd,ganeshan2017odd,Khain_2022,hosaka2023hydrodynamics,hosaka2021hydrodynamic,Hosaka2023,PhysRevE.90.063005,olvera,PhysRevLett.122.154501}.
\newline 
\mj{When an obstacle moves with respect to the fluid that it is embedded in, it is possible that it experiences a force that is orthogonal to this movement. Such an orthogonal force is called lift force. The inherent parity-breaking nature of odd viscosity suggests that in its presence, lift force can arise even when the geometry of the fluid system is rotationally symmetric.} For odd viscous flow past spheres, where the flow is along the preferred plane of odd viscosity \cite{PhysRevLett.127.048001,Khain_2022,reynolds2023dimensional}, lift is indeed found to be nonvanishing \cite{hosaka2023lorentz,everts2023dissipative,khain2023trading}. For two-dimensional fluid systems such as flow past infinite cylinders, it was found that this does not generally happen, because the integral that gives the odd viscous force on an obstacle is a closed contour  \cite{ganeshan2017odd}. Furthermore, \mj{when the flow past the cylinder is incompressible}, 
odd viscosity can be absorbed into the pressure so that odd viscosity effectively drops out of the Navier-Stokes equation. \mj{Lastly, no-slip boundary conditions prevent odd viscosity from affecting the fluid profile through stress-dependent slip effects. Because of this, odd viscosity is entirely unobservable and lift force vanishes in symmetric geometries for flow past cylinders if both incompressibility and no-slip boundary conditions hold}.
\newline 
\mj{For fluids that are compressible, it was found that lift force can arise \cite{hosaka2021nonreciprocal,lier2022lift}.} In a similar way, one can ask what happens when the no-slip boundary condition is lifted. This is not merely a theoretical exercise, because although the assumption of no-slip boundary condition is common, many fluid systems violate this condition and instead are characterized by obstacles that display a slip length that can be of micrometer scale \cite{lauga2005microfluidics,10.1063/1.1432696,Pit2000,Léger1999}. In this work, we consider the effect of a nonzero slip length at the boundary of a two-dimensional disk or three-dimensional cylinder on odd viscous flow and lift force. 
\newline 
The structure of this work is as follows: In Section~\ref{eq:oddincopmressiblefluids}, we repeat the derivation of Ref.~\cite{ganeshan2017odd}, stating that incompressible odd fluids with no-slip boundary conditions exhibit a flow independent of odd viscosity, and the absence of lift force. In Section~\ref{eq:lorentzreciprocaltheorem}, we show how through the odd generalization of the Lorentz reciprocal theorem \cite{hosaka2023lorentz}, we can obtain a simple formula for the cylinder forces at first order in slip length, which applies for general Stokesian fluids. It tells us that slip-induced lift vanishes linearly in slip length. For cylinder flow, Stokesian fluids suffer from the Stokes paradox \cite{veysey2007simple}, indicating that inertial contributions persist far from the cylinders unless an additional element is introduced to curtail this far-field flow. In Section~\ref{sec:momrelstokes}, we exactly solve for the cylinder flow when the Stokes paradox is circumvented by considering a thin sheet such as a membrane, which relaxes momentum to a neighboring fluid with much lower viscosity. We obtain a formula for the cylinder forces which depends non-perturbatively on slip length. We find that lift force does arise at quadratic order in slip length. Similarly, we compute the nonvanishing lift coefficient at quadratic order in slip length for the case of Stokes flow with a finite system size in Section~\ref{sec:finitesystemsize}. In Section~\ref{sec:oseenequaiton},we go beyond the Stokes approximation by solving the Oseen equation in a harmonic expansion. Imposing the boundary conditions up to the first harmonic leads to six equations which can be solved analytically to obtain the slip-induced odd viscous Oseen flow given in Fig.~\ref{fig:enter-label2981}. We also compute lift force and find that at leading order it also appears quadratically in slip length.

\section{No-slip boundary conditions}
\label{eq:oddincopmressiblefluids}   
We first describe the fluid equations and explain the result of Ref.~\cite{ganeshan2017odd} that in general, odd viscosity cannot fluids cannot produce a lift force on cylinders in the case of incompressiblity and no-slip boundary conditions. Because we exclusively consider cylinder flow, the third dimension plays no role, and we thus start from the two-dimensional Navier-Stokes equation for steady flow
\begin{align} \label{eq:navierstokesequation}
   \rho_0  v_j \partial_j v_i   =  \partial_j \sigma_{ij} ~~ ,  ~~ \partial_i v_j  =0 ~~ , 
\end{align}
where $\rho_0 $ is the density, $v_i$ is the two-dimensional fluid velocity and a summation over repeated indices is implied. $\sigma_{ij}$ is the stress tensor, which is constituted by
\begin{align} \label{eq:stressthing}
    \sigma_{ij}  =- p \delta_{ij}   +   \eta_s ( \partial_{i} v_{j} + \partial_{j} v_{i}) + \eta_o ( \partial_i v^*_j+ \partial^*_i v_j )  ~~ .
\end{align}
Here, $p$ is the pressure, and $\eta_s$ and $\eta_o$ are the shear and odd viscosity. We furthermore used the notation $a^{*}_i = \varepsilon_{ij } a_j $ for a general vector $a_j$. As was noted in Ref.~\cite{ganeshan2017odd}, incompressibility allows one to absorb the odd stress contribution into the pressure so that Eq.~\eqref{eq:navierstokesequation} turns into
\begin{align} \label{eq:navierstokesequation1}
   \rho_0  v_j \partial_j v_i   =  -  \partial_i \tilde p  + \eta_s \Delta v_i ~~ ,   
\end{align}
where we have introduced the modified pressure $\tilde p = p - \eta_o \partial_j v^*_j $ and $\Delta = \partial^2 $. For incompressible fluids, the only role of pressure is to guarantee the satisfaction of incompressibility, and this role can now be played by the modified pressure, so that odd viscosity drops out of the fluid equations. When the boundary conditions do not depend on stress, as is the case for no-slip boundary conditions, it thus follows that odd viscosity cannot affect the fluid profile. \newline 
We then consider a fluid system where a cylinder with no-slip boundary conditions is placed in a fluid described by Eq.~\eqref{eq:navierstokesequation1} and which moves with a velocity $U_i$. Even though the effect of odd viscosity is not observable in the fluid, there may still be an odd lift force for this fluid system because the force that this fluid exerts on the boundary is determined by the stress tensor which contains an odd viscous contribution. To check this, we first rewrite Eq.~\eqref{eq:stressthing} as
\begin{align} \label{eq:stress110909}
    \sigma_{ij}  =- \tilde  p \delta_{ij}   +   \eta_s (   \partial_{i} v_{j} +\partial_{j} v_{i}  )  + 2 \eta_o \partial^*_i v_j   ~~ .
\end{align}
Only the rightmost term in Eq.~\eqref{eq:stress110909} can produce lift force as it depends on odd viscosity, but this contribution always vanishes \cite{ganeshan2017odd}. Specifically, using that only the $\eta_o$-term in Eq.~\eqref{eq:stress110909} can produce a lift force $F_L$, closedness of the contour $\Gamma$ around the cylinder boundary leads to
\begin{align} \label{eq:nolift}
   F_L    =& \hat{U}^{*}_j    \oint_{\Gamma} ds \,   n_i \sigma_{i  j } = 2 \eta_o \hat{U}^{*}_j    \oint_{\Gamma} ds \,   n_i   \partial^*_i v_j =0   ~~ ,
      \end{align}
where $n_i$ is the normal vector for the cylinder boundary. Note that Eq.~\eqref{eq:nolift} does not merely apply to circular cylinders, although that is the case that is considered in the rest of this work.

\section{Lorentz reciprocal theorem}
\label{eq:lorentzreciprocaltheorem}
\begin{figure}[t]
    \centering
    \includegraphics[width=1\linewidth]{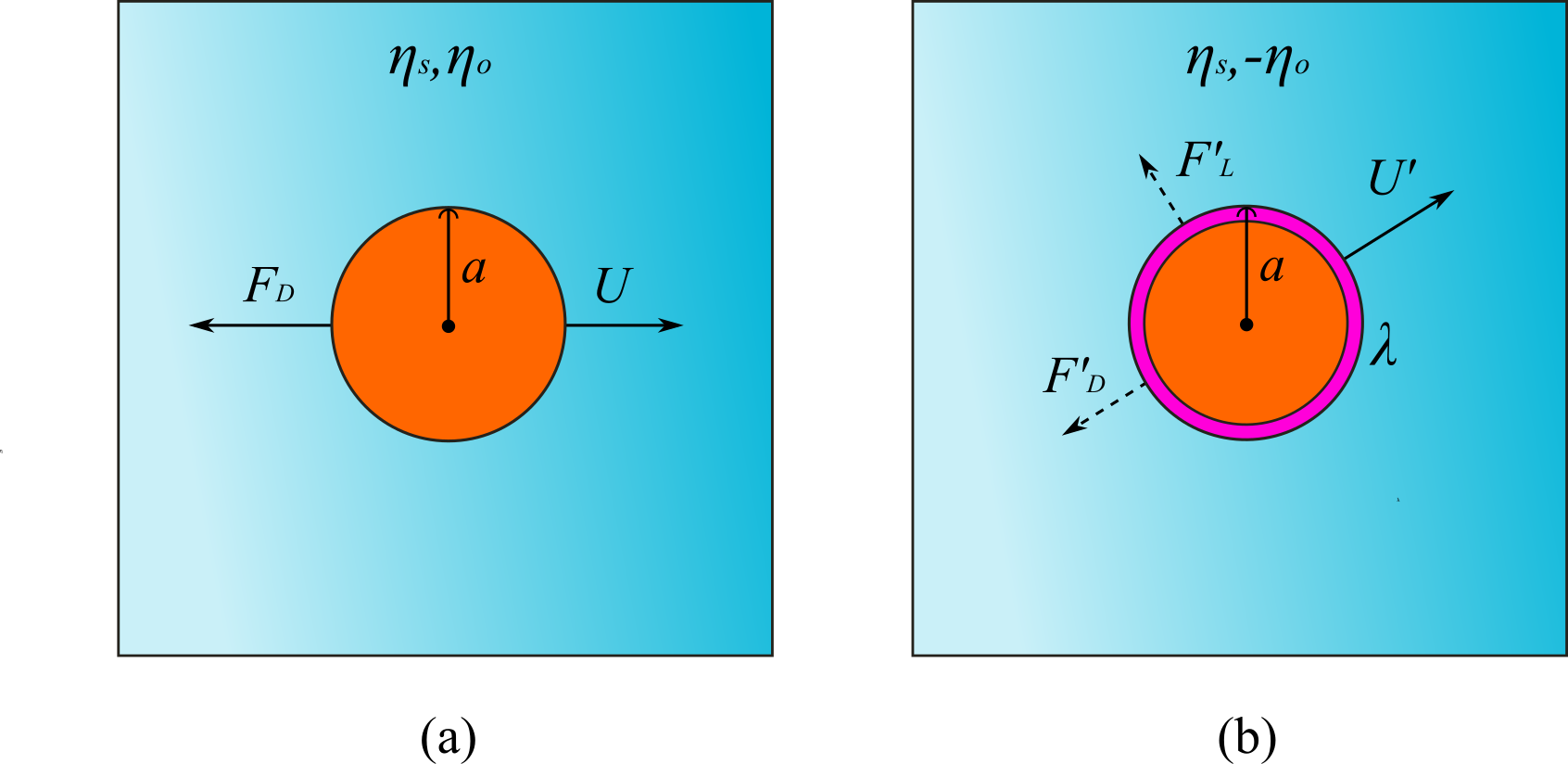}
    \caption{Schematic picture of two fluid systems that are being connected through the Lorentz reciprocal theorem generalized for odd viscosity. We connect (a) incompressible flow past a cylinder moving with velocity $U_i$ with no-slip boundary conditions giving rise to only drag and (b) incompressible flow past a cylinder moving with velocity $U^{\prime}_i$, which, to probe odd phenomena, should not generally be taken parallel to $U_i$. Because of the slip boundary conditions represented by the magenta strip, the second fluid system is not subject to the no-lift argument of Eq.~\eqref{eq:nolift} \cite{ganeshan2017odd}.}
    \label{fig:enter-label298}
\end{figure}

We now show that we can use the odd generalization of the Lorentz reciprocal theorem \cite{hosaka2023lorentz} to prove that, for incompressible odd Stokes flow past a cylinder, lift force vanishes not only at vanishing slip length but also at linear order in slip length. Stokesian fluids are characterized by the Stokes approximation, which assumes that the inertial contribution in Eq.~\eqref{eq:navierstokesequation1} is negligible, so that the fluid profile can be accurately described by
\begin{align} \label{eq:stokesequation}
  \partial_j \sigma_{ij}  =0 ~~ ,  ~~ \partial_i v_i   =0 ~~ , 
\end{align}
We consider two fluid systems for which Eq.~\eqref{eq:stokesequation} applies which will be connected through the Lorentz reciprocal theorem. The first fluid system is one where the no-slip boundary conditions hold and the fluid profile is thus completely even and the only force on the cylinder is drag force. The second fluid system we consider is identical except for three things:
\begin{enumerate}
\item We consider for the second fluid system a cylinder velocity $U^{\prime }_i$ which is not necessarily parallel to $U_i$, as a parallel $U^{\prime }_i$ would not allow one to extract lift force using the Lorentz reciprocal theorem. 
    \item For the second fluid system, we impose slip boundary conditions \cite{masoud2019reciprocal,legendre_lauga_magnaudet_2009}. Working in the frame where the fluid velocity vanishes far away from the cylinder, we have at the cylinder interface
\begin{align}  \label{eq:slipboundarycondition}
    v^{\prime \text{S}}_i \big|_{r= a } = U^{\prime   }_i + \frac{\lambda  }{\eta_s }  ( \delta_{ij}  - n_i n_j  ) n_k \sigma^{\prime \text{S}}_{ k j  }   ~~ , 
\end{align}
where $r = \sqrt{x^2 +y^2 }$, $a$ is the cylinder radius and $\lambda$ is the slip length. $\sigma^{\prime \text{S}}_{ k j  } $ is the stress corresponding to the fluid profile with slip. 
\item In order for the Lorentz reciprocal theorem to work for odd fluids, we require that odd viscosity of the second fluid system is given by $\eta^{\prime}_o =  - \eta_o$ \cite{hosaka2023lorentz}. 
\end{enumerate}
The two distinct fluid systems are schematically shown in Fig.~\ref{fig:enter-label298}. Because both fluid systems obey the Stokes equation and the first system fluid system obeys no-slip boundary conditions, there is the reciprocal relation \cite{lorentzoriginal}
\begin{align} \label{lorentzreciprocaltheoremapply}
  \oint_{\Gamma} d s \,  n_i  \sigma^{\prime \text{S} }_{ij} U_j = \oint_{\Gamma} d s \,  n_i  \sigma_{ij} v^{\prime \text{S} }_j   ~~ . 
\end{align}
We plug in Eq.~\eqref{eq:slipboundarycondition} to find 
\begin{align} \label{eq:thingggg}
 F^{\prime \text{S}}_i U_i   =   F_i U^{\prime  }_i   +  \frac{\lambda}{\eta_s }  \oint_{\Gamma} ds f^{  \prime \text{S} }_i ( \delta_{ij}  - n_i n_j  ) f_j   ~~  , 
\end{align}
where we introduced the force densities $f_j = n_i \sigma_{ij}$ and $f^{\prime \text{S} }_j = n_i \sigma^{\prime \text{S}}_{ij}$ and we defined the total forces $F_i =  \oint_{\Gamma} d s f_i $ and $F^{\prime \text{S}}_i =  \oint_{\Gamma} d s f^{\prime \text{S}}_i $. We then assume $\hat{ \lambda  } = \lambda a^{-1}$ to be small, so that we can introduce 
\begin{align}
     v^{\prime }_i =  v^{\prime \text{S}}_i  +  \mathcal{O}  ( \hat{ \lambda  }  ) ~~  ,
    \end{align}
where $  v^{\prime}_i$ is a fluid profile that obeys no-slip boundary conditions but still has the same far-field velocity $U^{\prime}_i$ and odd viscosity $\eta_o^{\prime}$. This fluid profile has a corresponding stress $\sigma^{\prime}_{ij}$ and force $f^{\prime}_i$. Eq.~\eqref{eq:thingggg} can then be expanded as \cite{masoud2019reciprocal}
\begin{align} \label{eq:masterformula}
 F^{\prime \text{S} }_i U_i   =   F_i U^{\prime }_i   +  \frac{\lambda }{\eta_s } \oint_{\Gamma} ds f^{ \prime }_i ( \delta_{ij}  - n_i n_j  ) f_j   + \mathcal{O} ( \hat \lambda^2 )   ~~  .  
\end{align}
Note that the modified pressure $\tilde p$ does not contribute to either slip-induced drag force or lift force. We now consider two cases, namely the case where $U_i = U^{\prime}_i$ and where $U_i = U^{\prime * }_i$. These cases allow one to extract the slip-generalized drag and lift coefficients $C_D$ and $C_L$ which are defined as
\begin{align}
     F^{\prime \text{S} }_i  =  \left( -C_D \delta_{ij} + C_L  \epsilon_{ij} \right) U^{\prime}_j ~~ . 
\end{align}
  Taking $ U^{\prime}_i =   U_i$ for Eq.~\eqref{eq:masterformula} yields
\begin{align}
\begin{split}
    C_D    & =     C^{(0)}_D +   \hat \lambda C^{(1)}_D + \mathcal{O} ( \hat \lambda^2 )
    \end{split}
    \end{align}
where 
\begin{subequations} \label{eq:theinitialequations}
   \begin{align}
    C^{(1)}_D   = -   \frac{a  }{\eta_s |U|^2 } \oint_{\Gamma} ds f_i ( \delta_{ij}  - n_i n_j  ) f_j    ~~ ,   
   \end{align}
   and $  C^{(0)}_D$ is the drag force corresponding to the fluid system with no-slip boundary condition. Taking $   U^{\prime }_i =    U^{*}_i $  for Eq.~\eqref{eq:masterformula} yields $  C_L^{(0)}  =0 $ and
\begin{align} \label{eq:liftforceformula}
    C_L^{(1)}  = &    \frac{a  }{\eta_s |U|^2 } \oint_{\Gamma} ds f^{ * }_i ( \delta_{ij}  - n_i n_j  ) f_j    ~~ , 
\end{align}    
\end{subequations}
For fluid systems described by incompressible Stokes flow and no-slip boundary conditions, analytical expressions for $v_i$ and $ C^{(0)}_D $ are often readily available, so that these solutions can be plugged into Eq.~\eqref{eq:theinitialequations} to obtain slip-induced forces. Furthermore, the rotational symmetry and linearity of the Stokes equation makes it so that fluid profiles for a geometry with cylindrical symmetry and cylinder velocity orthogonal to the cylinder axis can generally be written as $v_i = \partial^{*}_i \psi $ with \cite{rosenhead1963laminar,dolfo2020stokes}
\begin{align} \label{eq:psidefinition}
    \psi =  -  \frac{  U^{*}_i x_i   }{r}  g(r )  ~~ , 
\end{align}
where $g(r ) $ is some function that can be obtained by solving the Stokes equation. When Eq.~\eqref{eq:psidefinition} holds, Eq.~\eqref{eq:liftforceformula} turns into 
    \begin{align} \label{eq:Cformulag}
    C^{(1)}_L  =     \frac{  4 \pi    \eta_o  }{a^2}  \left( g -  a \dot{g} \right) \left(a^2 \ddot{g}-a \dot{g} +g\right)    ~~ , 
\end{align} 
where $g$ is shorthand for $g(a)$. Furthermore, $\dot{g} = \partial_r g(r)|_{r=a}$ and $\ddot{g} = \partial^2_r g(r)|_{r=a}$. The no-slip boundary condition implies $g = a \dot{g}$, so that Eq.~\eqref{eq:Cformulag} reduces to
\begin{subequations} \label{eq:finalexpressionsss}
\begin{align} \label{eq:vanishingliftforce}
      C^{(1)}_L  = 0 ~~ . 
\end{align}
Similarly, $C_D^{(1)}$ is given by the simple expression
\begin{align} \label{eq:Cformuladrag}
\begin{split}
    C^{(1)}_D   &   =  -  \pi a^2   \eta_s    \ddot{g}^2       ~~  . 
\end{split}
 \end{align} 
\end{subequations}
Note that $C^{(1)}_D $ is non-positive which can be understood by considering that slip will always serve to ease flow past an obstacle and thus lower drag force \cite{masoud2019reciprocal}. Eq.~\eqref{eq:vanishingliftforce} does not mean that lift force remains zero when turning on finite slip length, but only that for fluids for which Eq.~\eqref{lorentzreciprocaltheoremapply} holds and one has a fluid velocity of the form of Eq.~\eqref{eq:psidefinition}, there is no contribution linear in slip length. In the following sections we consider specific fluid systems where we indeed find a nonvanishing lift force which at leading order is quadratic in slip length.

\section{Stokes flow in a thin sheet}
\label{sec:momrelstokes}
As an example of a fluid system with an obstacle for which the Stokes approximation holds, we consider the case that the fluid lies in a thin sheet which is connected to a bulk fluid with a much lower viscosity to which it relaxes momentum \cite{seki1993brownian,saffman1975brownian,saffman1976brownian}. The thin sheet should be seen as two-dimensional and the obstacle that lies inside of the thin sheet as a disk. We assume that the relaxational effects to the bulk fluid dominate over the inertial effects\footnote{ Interestingly, such an overdamped setup is mathematically consistent with that of a droplet of \quotes{spinner fluid} lying on a glass plate ~\cite{soni2019odd}. The spinner fluid is a suspension of magnetically driven chiral colloids. The spinner fluid is the first experimental system for which odd viscosity was measured at the micrometer scale.}. When there is a leak of momentum to a three-dimensional bulk, the Stokes equation modifies to the Stokes-Brinkman equation \cite{Brinkman1949}, which is given by
\begin{align} \label{eq:lalalala}
      \partial_j  \sigma_{ij}   =   \frac{ \rho_0 }{\tau }  v_i      ~~ , 
\end{align}
where $\tau$ is the momentum relaxation time and $\rho_0$ is the density. This relaxation term will allow for the Stokes paradox to be circumvented. Note that this modification of the Stokes equation does not invalidate formula based on the  Lorentz reciprocal theorem, as this momentum relaxation term would cancel out in Eq.~\eqref{lorentzreciprocaltheoremapply}. The solutions to Eq.~\eqref{eq:lalalala} can be decomposed into even and odd ones as
\begin{align} \label{eq:simpleansatz}
       \psi =  -\frac{  x_i   }{r}  \left[ U^{*}_i g_e (r )  +  U_i   g_o(r ) \right]   ~~ . 
\end{align}
To obtain the solutions, we take the curl of Eq.~\eqref{eq:lalalala} to find
\begin{align} \label{eq:lalalala1212}
  \Delta   \left[   \Delta   
 -   \kappa^2  \right] \psi   =  0       ~~ , ~~ \kappa^2  = \rho_0 / (\tau \eta_s )  ~~ .    
\end{align}
Solving Eq.~\eqref{eq:lalalala1212} and requiring convergence to zero at infinity leads to
    \begin{align} \label{eq:solutiobnszzz}
     g_{e,o} (r )   =  A_{e,o}  a^2 r^{-1}    +   B_{e,o}   a  K_1 (\kappa r )  ~~ , 
\end{align}
where $K_n ( x  )$ is the $n$th modified Bessel function of the second kind. We impose the slip boundary conditions 
\begin{align}  \label{eq:slipboundarycondition1}
    v_i  \big|_{r= a }  = U_i + \frac{\lambda  }{\eta_s }  ( \delta_{ij}  - n_i n_j  ) n_k \sigma_{ k j  }   ~~ , 
\end{align}
which leads to constraint equations for $A_{e,o}$ and $B_{e,o}$ given by
\begin{align}
     M_{4 \times 4 }  \begin{bmatrix}
         A_e    &  
               B_e    & 
             A_o   & 
               B_o   
\end{bmatrix}^T =    \begin{bmatrix}
         1    &  
           -    1    & 
             0     & 
               0  
\end{bmatrix}^T ~~ , 
\end{align}
with 
   \begin{align}
M_{4 \times 4 }  = \begin{bmatrix}
 1 & K_1(\hat \kappa) & 0 & 0 \\
 4 \hat{\lambda}+1 & \Theta (\hat \kappa , \hat \lambda )  & -4 \gamma_o  \hat{\lambda} &  \gamma_o \hat \lambda 
 \Psi (\hat \kappa ) 
  \\
 0 & 0 & 1 & K_1 (\hat \kappa) \\
 -4 \gamma_o  \hat{\lambda} & \gamma_o  \hat \lambda \Psi (\hat \kappa ) & -4 \hat{\lambda}-1 & -\Theta (\hat \kappa , \hat \lambda ) 
\end{bmatrix}   ~~  , 
   \end{align}
where $\hat \kappa = \kappa a $, $\gamma_o = \eta_o / \eta_s$ and
    \begin{align}
    \begin{split}
        \Theta  (\hat \kappa , \hat \lambda )  &  =  \left(\hat{\lambda} \hat \kappa^2+4 \hat{\lambda} +1\right) K_1 (\hat \kappa) + \hat \kappa(2 \hat{\lambda}+1) K_0 ( \hat \kappa) ~~ ,   \\ 
\Psi  (\hat \kappa )  &  = -2   \hat \kappa  K_2 (\hat \kappa)~~ .  
    \end{split}
\end{align}
Having found the coefficients of Eq.~\eqref{eq:solutiobnszzz}, we compute the cylinder forces 
\begin{subequations} \label{eq:forcefomulae}
    \begin{align}
    F_D &  = a  \int_0^{2 \pi } d \theta  \left[   \sigma_{r r } \cos(\theta )  -    \sigma_{r \theta }  \sin(\theta) \right] ~~ ,   \\ 
 F_L    & = a  \int_0^{2 \pi } d \theta  \left[   \sigma_{r r } \sin(\theta )   +     \sigma_{r \theta }  \cos(\theta) \right] ~~ , 
\end{align}
\end{subequations}
where 
\begin{subequations}
    \begin{align}
    \begin{split}
            \sigma_{r r }   &  =  - \tilde p  + 2 \eta_s  \partial_r v_r   + 2 \eta_o  r^{-1}   \left( \partial_{\theta } v_r  - v_{\theta } \right)    ~~ , 
      \end{split}
\\ 
    \begin{split}
            \sigma_{r \theta  }  &  = \eta_s \left[ r^{-1}   \partial_\theta   v_{r  }  + r \partial_r  \left( v_{\theta } r^{-1 } \right) 
 \right] 
 -  2 \eta_o   \partial_{r } v_{r } ~~ . 
     \end{split}
\end{align}
\end{subequations}
$ \tilde p$ can be obtained up to an unimportant constant $c$ by $\theta $-integrating the $\theta$-component of the Stokes equation, i.e. 
\begin{align}
\begin{split}
        &   \tilde p - c   =  \\  & \eta_s  \int d \theta \left[  r^{-1}\partial_r (r v_\theta
    ) + \frac{1}{r^2 } (  \partial_{\theta}^2   + 2  \partial_{\theta }   )v_r -  \left( r^{-2} +\kappa^2 \right)   v_\theta   \right] .  
\end{split}
\end{align}
The formulae for the drag and lift coefficients are then given by
    \begin{align} \label{eq:formulae}
   C_{D , L} & = \frac{\pi  \eta_s   }{a}   \left[ a  \left(\kappa ^2 a^2+3\right) \dot{g}_{e,o}    - 3 g_{e,o}   - a^3 \dddot{g}_{e,o} \right]   ~~ , \end{align}
   where, $\dddot{g} = \partial^3_r g(r)|_{r=a}$. One can see that are no odd viscosity terms in Eq.~\eqref{eq:formulae}, which is due to the argument in Eq.~\eqref{eq:nolift} that ruled out lift for fluid flows that do not see odd viscosity. Instead, odd viscosity only enters indirectly in the drag and lift formulae through $g_{e,o}$. Plugging the solutions for $g_{e,o}$ into Eq.~\eqref{eq:formulae}, we obtain for the drag coefficient at leading and subleading order in slip length
\begin{subequations}
\begin{align}
   C_D^{(0)}     & =  \pi  \eta_s  \hat \kappa   \left[\hat \kappa   + \frac{4 K_1(\hat \kappa  )}{K_0( \hat \kappa  )}\right]  ~~ ,    \\ 
   C_D^{(1)}     & =   -  \frac{4 \pi  \eta_s \hat  \kappa ^2 K^2_1( \hat \kappa  )}{  K^2_0( \hat \kappa  )   }  \label{eq:firstorder}         ~~ . 
   \end{align}
\end{subequations}
As was predicted using the Lorentz reciprocal theorem, we find $  C_L^{(1)}  =0$. However, at second order in slip length there is lift given by
    \begin{align} 
    \begin{split}
            \label{eq:finalsolution}
  &  C^{(2)}_L     =   \\  &   \frac{32 \pi   \eta_o \hat  \kappa ^4   K_1^2 ( \hat \kappa  )}{  \hat \kappa^2 \left[ 3 K_0^2 ( \hat \kappa  )+   K_2^2( \hat \kappa  )  \right]  -4 K_1 ( \hat \kappa  )  \left[  
 \hat \kappa  K_0 ( \hat \kappa  )  +  K_1 ( \hat \kappa  )  \right]  }   ~~ .
     \end{split}
\end{align}
We learn from Eq.~\eqref{eq:finalsolution} that although lift force vanishes entirely with no-slip boundary conditions and the generalized Lorentz reciprocal theorem tells us that the lift force contribution linear in slip length vanishes, nonvanishing contributions to lift force do appear at quadratic order in slip length. Note that although we did not expand in small odd viscosity, Eq.~\eqref{eq:finalsolution} does not display any non-linear odd viscous contributions as odd viscosity can only enter through the slip boundary condition, making it so that any result for drag or lift we find is still indirectly suppressed in odd viscosity. The results of this section can be mapped to the problem of drag and lift for oscillating cylinders in a fluid which is otherwise stationary. When a cylinder is oscillating with frequency $\omega$, the fluid motion will display the same oscillation provided that the equation which the fluid obeys is linear. Using complex notation, the fluid velocity is therefore $\sim \exp(i \omega t )$, where $i = \sqrt{-1}$, and the Navier-Stokes equation reduces to \cite{williams1972oscillating,Hussey1967,rosenhead1963laminar,landau2011fluid,PhysRevE.63.041510}
\begin{align} \label{eq:lalalala12}
      \partial_j  \sigma_{ij}   =  i  \omega   \rho_0   v_i      ~~ . 
\end{align}
Comparing Eq.~\eqref{eq:lalalala12} to \eqref{eq:lalalala}, one finds that all one must do to obtain an expression for drag and lift of oscillating cylinders in the frequency domain is replace $\tau^{-1}$ by $i \omega $.
\section{Finite system size}
\label{sec:finitesystemsize}
 Another fluid system for which the drag and lift coefficient formulae Eq.~\eqref{eq:finalexpressionsss} could be applied is that of Stokes flow confined by an outer cylinder \cite{dolfo2020stokes,lucas2017stokes}, provided that one only considers slip for the inner cylinder and assumes no-slip boundary conditions for the outer cylinder that sets the system size. The curl of the Stokes equation is given by $   \Delta^2 \phi =0 $ so that, using again Eq.~\eqref{eq:simpleansatz}, we can solve the Stokes equation with the ansatz    \begin{align} \label{eq:solutiobnszzz11}
     g_{e,o} (r )   =  A_{e,o}  \frac{a^2}{r}   +   B_{e,o}    r   +   C_{e,o}   r   \log(\frac{r}{a} )     +   D_{e,o}   \frac{r^3}{a^2}~~ . 
\end{align}
 One can acquire the expressions for the coefficients of Eq.~\eqref{eq:solutiobnszzz11} by imposing the boundary conditions of Eq.~\eqref{eq:slipboundarycondition1} as well as the boundary condition of the outer cylinder at radius $b$ given by
\begin{align}  \label{eq:slipboundarycondition2}
    v_i  \big|_{r= b }  =0  ~~ . 
\end{align}
We then obtain the drag coefficients
\begin{subequations}
\begin{align}
   C_D^{(0)}     & = \frac{4 \pi  \eta_s  \left(a^2+b^2\right)}{\left(a^2+b^2\right) \log \left(\frac{b}{a}\right)+a^2-b^2 }  ~~ ,    \\ 
   C_D^{(1)}     & =   -  \frac{4 \pi  \eta_s  \left(a^2-b^2\right)^2}{ \left[\left(a^2+b^2\right) \log \left(\frac{b}{a}\right)+a^2-b^2\right]^2}  \label{eq:firstorder12}         ~~ , 
   \end{align}
\end{subequations}
as well as leading order lift coefficient
   \begin{align} 
    \begin{split}
            \label{eq:finalsolution1}
  &  C^{(2)}_L     =  \frac{8 \pi  \eta_o \left(a^2-b^2\right)^2}{\left[\left(a^2+b^2\right) \log \left(\frac{b}{a}\right)+a^2-b^2\right]^2} ~~ .
     \end{split}
\end{align}

\section{Oseen flow}
\label{sec:oseenequaiton}
\begin{figure}[t]
    \centering \includegraphics[width=1\linewidth]{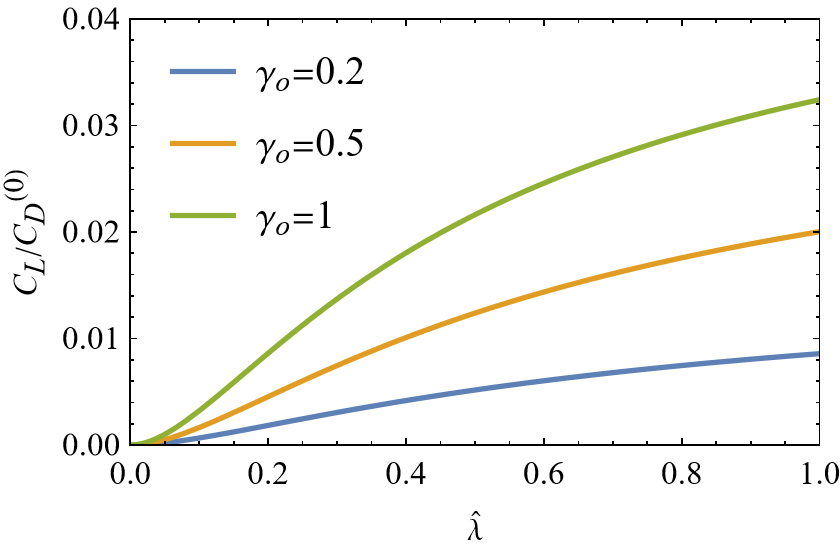}
    \caption{Plot of Oseen lift coefficient as a function of slip length. We took $\hat k = 0.2$.}
    \label{fig:enter-label29811}
\end{figure}

In the previous sections we have relied on the Stokes approximation. We now wish to explore the effect of slip and odd viscosity for the case where the Stokes approximation is not used, i.e. where the inertial term in Eq.~\eqref{eq:navierstokesequation} is taken into account. This inertial term makes sure that the Stokes paradox no longer arises. Furthermore, we can no longer use the drag and lift coefficient formulae of Eq.~\eqref{eq:finalexpressionsss}, as they follow from a result which is only valid for Stokes flow. In Ref.~\cite{LI2014211}, the drag force for even cylinder flow with inertia is computed by generalizing the computation by Kaplun \cite{kaplun1957low} to one with slip boundary conditions. This computation uses asymptotic matching \cite{proudman1957expansions,veysey2007simple,chester_breach_proudman_1969,vandyke1975perturbation}, which is a method where the fluid equations are solved near to and away from the cylinder by systematically matching the solution in the transition region. In App.~\ref{app:oseendragslip1}, we rederive this result with the method described below. This method is essentially equivalent to the near-cylinder harmonic expansion performed by Lamb \cite{lamb1932hydrodynamics}. However, to generalize the computation of Oseen flow to the case where there is a nonvanishing odd viscosity is not straightforward, and therefore it helps to perform a more systematic harmonic expansion for obtaining the near-cylinder flow similar to how it is done in the computation of the even Oseen flow by Tomotika and Aoi \cite{tomotikaaa}. \newline 
We start by performing the Oseen approximation \cite{oseen1910uber,lamb1932hydrodynamics} on Eq.~\eqref{eq:navierstokesequation}. For this, we move to the frame that is co-moving with the cylinder, i.e. it holds that $\lim_{r \rightarrow \infty} v_i = -U_i $. We then define a fluid velocity $ v_i^{\prime} = U_i + v_i  $. As $v_i^{\prime}$ vanishes for $r \rightarrow \infty$, $v_i^{\prime}$ can be viewed as a small correction to $U_i$ when one is sufficiently far from the cylinder. We thus linearize Eq.~\eqref{eq:navierstokesequation}, which leads to
\begin{align} \label{eq:oseenequation1}
   2 \eta_s  k  \partial_x  v^{\prime }_i  
 -\partial_j  \tilde p   +  \eta_s    \Delta v^{\prime }_i   =0  ~~ , ~~ \partial_i v^{\prime }_i  =0 ~~ ,   
\end{align}
with $k   = \frac{1}{2} \rho_0 U  / \eta_s $. We can solve Eq.~\eqref{eq:oseenequation1} for
\begin{subequations}
    \begin{align}
    \tilde  p     &  = - 2 k \eta_s  \partial_x \phi\\ 
          v^{\prime }_x   & =   -  \partial_x \phi  +   \frac{1}{2 k } \partial_x  \chi    +  \chi   ~~ , \\ 
          v^{\prime }_y   &  =   -  \partial_y \phi  +    \frac{1}{2 k   } \partial_y   \chi    ~~  , 
\end{align}
\end{subequations}
where $\phi$ and $\chi$ are functions that satisfy the equations
\begin{align} \label{eq:differentialequations}
    \Delta \phi =0 ~~ , ~~  \left( \Delta   +  2 k \partial_x  \right) \chi =0  ~~ . 
\end{align}
The families of solutions to Eq.~\eqref{eq:differentialequations} are given by the even functions
\begin{subequations} \label{eq:evensolutions}
    \begin{align}
    \phi_e   &    = a U A_0  \log (r )  + U \sum_{n =1}    A_n \frac{ a^{n +1}   }{ r^n} \cos(n \theta ) ~~ ,   \\ 
        \chi_e    &  =  U e^{  - k r \cos(\theta ) }   \sum_{m=0}    B_m K_m (k r )  \cos(m \theta ) ~~ , 
\end{align}
\end{subequations}
as well as the odd functions
\begin{subequations} \label{eq:oddsolutions}
    \begin{align}
    \phi_o    &  =   U \sum_{n=1} \tilde A_n  \frac{ a^{n +1}     }{ r^n}  \sin( n \theta) ~~ ,   \\ 
        \chi_o   &  =  U e^{ - k r \cos(\theta ) }  \sum_{m=1}   \tilde B_{m} K_{m} (k r ) \sin(m \theta )  ~~  . 
\end{align}
\end{subequations}
In Ref.~\cite{tomotikaaa}, Tomotika and Aoi impose no-slip boundary conditions for all harmonics to obtain equations that completely fix $A_n$ and $B_n$ up to arbitrarily high orders,
after which they perform a truncation to obtain analytic expression for the fluid profile and drag. Due to the nature of the solutions which carry the constant $k$ inside the Bessel function, such a truncation of the harmonic expansion is simultaneously a truncation of a small Reynolds number expansion. Therefore, the obtained drag is found at leading order in Reynolds number to coincide with the result found by Kaplun using asymptotic matching \cite{kaplun1957low}. We avoid working out the slip boundary conditions for the infinite harmonic series and instead first perform the truncation. Specifically, we expand up to the first harmonic, which means that there are three boundary conditions that constrain the even flow and three boundary conditions that constrain the odd flow. It is then sensible to work with an ansatz that is composed of three even solutions and three odd solutions given in Eqs.~\eqref{eq:evensolutions} and~\eqref{eq:oddsolutions} respectively. 
\begin{figure*}[t]
    \centering \includegraphics[width=1.0\linewidth]{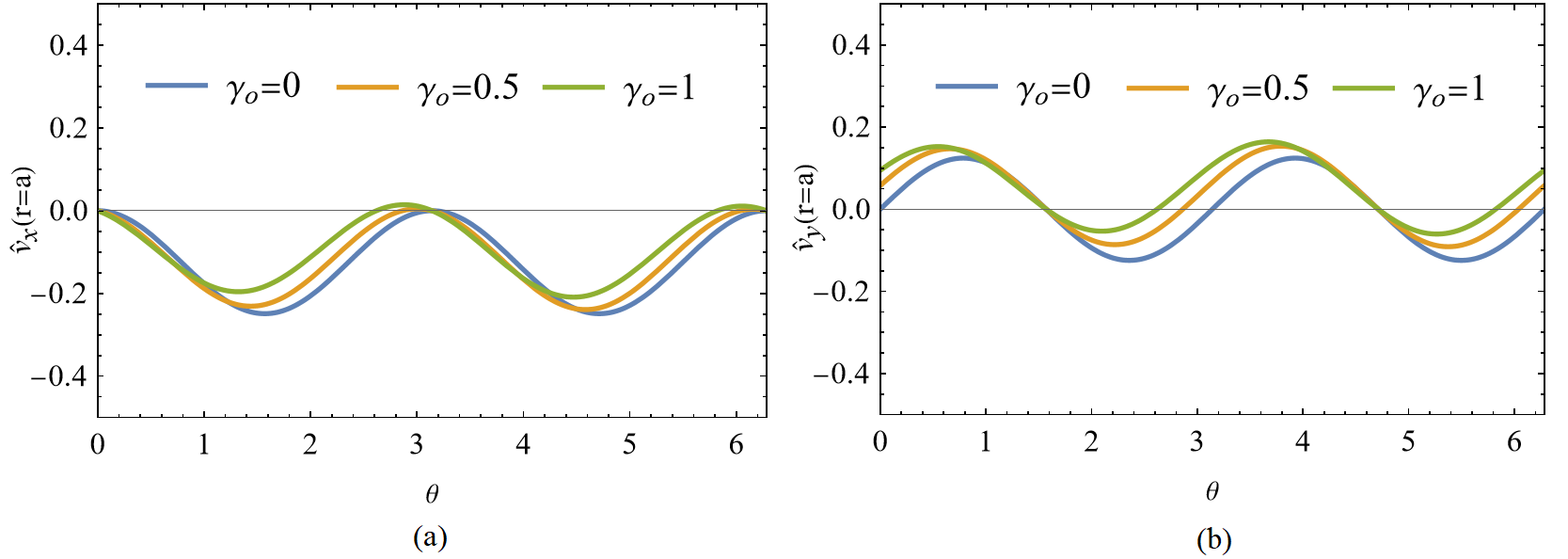}
    \caption{Picture of (a) the horizontal dimensionless slip velocity $\hat v_x (r=a)$ and (b) the vertical dimensionless slip velocity $\hat v_y (r=a)$. We took $\hat \lambda =1 $ and $\hat k =0.2 $.}
    \label{fig:enter-label29812}
\end{figure*}
For the even part of the ansatz, we make the default choice \cite{tomotikaaa,lamb1932hydrodynamics,PhysRevB.99.035430} which involves expanding Eq.~\eqref{eq:evensolutions} up to $n=1$ and $m=0$ \footnote{Choosing instead to expand up to $n=0$ and $m=1$ does not affect the Oseen flow qualitatively. At low Reynolds number, it also does not affect the flow quantitatively. See Appendix~\ref{app:oseendragslip1} where it is found that the contribution to Oseen drag with slip that is leading order for small Reynolds number is identical regardless of the choice of ansatz.}. For the odd solutions, one could either expand Eq.~\eqref{eq:oddsolutions} up to $n=2$ and $m=1$ or expand up to $n=1$ and $m=2$. Only the latter option can satisfy the boundary conditions as the $\tilde A_2$-solution does not affect the boundary conditions up to first order in a harmonic expansion. Having narrowed down the options for the ansatz to one, we formulate the constraints
\begin{align} \label{eq:constraints}
   & \left(M_{6 \times 6}^{(0)}   + \hat \lambda M_{6 \times 6}^{(1)}\right) \vec{V}   =   \begin{bmatrix}
     0    & 
            1    &  
            -     1   &  
      0    &  
             0   & 
               0   
\end{bmatrix}^T ~~ , 
\end{align}
where $\vec{V} =  \begin{bmatrix}
     A_0     &     B_0   &   A_1   &  \tilde A_1    &  \tilde B_1  &   \tilde B_2 
\end{bmatrix}^T$. Because the matrix $M^{(1)}_{6 \times 6}   $ is large, we decompose as $M^{(1)}_{6 \times 6}   = \begin{bmatrix}
    M^{(1,a)}_{6 \times 4}   & M^{(1,b)}_{6 \times 2}  
\end{bmatrix}$. The content of the matrices is given by
\begin{widetext}
\begin{subequations}
    \begin{align}  \label{eq:matricesoseen}
    M^{(0)}_{6 \times 6} &  = \begin{bmatrix} -1 & -\frac{1}{2 \hat{k}} & 0 & 0 & 0 & 0 \\
 0 & I_2 K_0+I_1 \left(\frac{K_0}{\hat{k}}+K_1\right) & 1 & 0 & 0 & 0 \\
 0 & -\frac{I_1 K_0}{\hat{k}} & 1 & 0 & 0 & 0 \\
 0 & 0 & 0 & 0 & -\frac{I_1 K_1}{\hat{k}} & \frac{2 I_2 K_2}{\hat{k}} \\
 0 & 0 & 0 & 1 & \frac{I_1 K_1-1}{\hat{k}^2} & \frac{2-2 I_2 K_2}{\hat{k}^2} \\
 0 & 0 & 0 & -1 & \frac{(I_1+2 \hat{k} I_2) K_1}{\hat{k}^2} & \frac{2 (5 I_2-2 \hat{k} I_1) K_2}{\hat{k}^2} 
\end{bmatrix} ~~ ,  \\ 
   M^{(1,a)}_{6 \times 4}   &   =  \begin{bmatrix}
     0 & 0 & 0 & 0  \\
 0 & 0 & 0 & 0  \\
 0 & 2 I_2 K_0 & 4 & - 4 \gamma_o \\
 2 \gamma_o & \frac{\gamma_o}{\hat{k}} & 0 & 0  \\
 0 & 0 & 0 & 0  \\
 0 & -2 \gamma_o (I_2 K_0+I_1 K_1) & -4 \gamma_o & - 4    \end{bmatrix} ~~ , \\ 
   M^{(1,b)}_{6 \times 2}   &   =  \begin{bmatrix}
     0 & 0 \\
  0 & 0 \\
  \frac{2 \gamma_o \left[ 3 \hat{k} I_0 K_1+I_1 (\hat{k} K_0-4 K_1) \right] }{\hat{k}^2} & -\frac{4 \gamma_o \left[\hat{k} I_0 \left(\left(\hat{k}^2-12\right) K_1-6 \hat{k} K_0\right)+I_1 \left(3 \hat{k} \left(\hat{k}^2+4\right) K_0+4 \left(\hat{k}^2+6\right) K_1\right)\right]}{\hat{k}^4} \\
 I_1 \left(-K_0-\frac{4 K_1}{\hat{k}}\right)+I_0 K_1 & \frac{2 \hat{k} I_0 \left[\left(\hat{k}^2+12\right) K_1+6 \hat{k} K_0\right]-2 I_1 \left[\hat{k} \left(\hat{k}^2+12\right) K_0+4 \left(\hat{k}^2+6\right) K_1\right]}{\hat{k}^3} \\
 0 & 0 \\
  \frac{2 \hat{k} I_0 (\hat{k} K_0+5 K_1)-2 I_1 \left[\left(\hat{k}^2+8\right) K_1+\hat{k} K_0\right]}{\hat{k}^2} & \frac{4 \hat{k} I_0 \left[\hat{k} \left(\hat{k}^2+18\right) K_0+4 \left(\hat{k}^2+9\right) K_1\right]-4 I_1 \left[4 \hat{k} \left(2 \hat{k}^2+9\right) K_0+\left(\hat{k}^4+20 \hat{k}^2+72\right) K_1\right]}{\hat{k}^4}  \end{bmatrix} ~~ , 
\end{align}    
\end{subequations}
\end{widetext}
where $I_n(x)$ is the $n$th modified Bessel function of the first kind. Furthermore, we used the definition $\hat k = k a $ and the Bessel functions in Eq.~\eqref{eq:matricesoseen} should be understood as having $\hat k $ in the argument. The solution to Eq.~\eqref{eq:constraints} leads to the fluid profile up to first order in a harmonic expansion given in Fig.~\ref{fig:enter-label2981}. At low Reynolds number, the slip-induced lift force shows the same qualitative properties as found for Stokes flow, namely $   C_L^{(1)}      = \mathcal{O} (\hat k ) $ and
    \begin{align}
        \begin{split}
                        C_L^{(2)}   &  =   \frac{96 \pi  \eta_o }{6 \left(2 \log \left(\frac{\hat k }{2}\right)+2 \gamma_{EM}\right)^2+2 \log \left(\frac{\hat k }{2}\right)+2  \gamma_{EM}-7}  \\  & 
            +  \mathcal{O} (\hat k ) ~~ , 
                    \end{split}
    \end{align}
    where $\gamma_{EM}$ is the Euler-Mascheroni constant. The $\hat \lambda$-dependence of the unexpanded lift coefficient is plotted in Fig.~\ref{fig:enter-label29811}. $  C_L^{(2)}  $ is nondimensionalized with the Oseen drag for zero, slip, whose expression is given by \cite{lamb1932hydrodynamics}
    \begin{align}
            C_D^{(0)}   &  =   \frac{4 \pi \eta_s}{ \frac{1}{2} - \log (\frac{\hat k}{2})- \gamma_{EM}  }    +  \mathcal{O} (\hat k ) ~~  . 
    \end{align}
    It can be seen in Fig.~\ref{fig:enter-label29811} that for the given values lift force is at least one order of magnitude smaller than drag, which does not mean that this lift force cannot be measured as it is a distinct physical effect which does not compete with drag force. \mj{To highlight the interplay between slip and odd viscosity, we also provide Fig.~\ref{fig:enter-label29812}, which shows the slip velocity around the boundary for different values of odd viscosity. Looking at the vertical slip velocity, we see that it is up-down asymmetric for nonzero odd viscosity. For vanishing odd viscosity, the slip velocity is up-down symmetric which is expected considering that the geometry of the fluid system is up-down symmetric.} Lastly, let us look at torque     $\tau$, which is given by
    \begin{align}
    T    = \oint_{\Gamma} ds \,  \varepsilon_{ij}  x_i n_k  \sigma_{kj}    ~~ .   
    \end{align}
    As shown in Ref.~\cite{ganeshan2017odd}, a nonvanishing torque cannot arise from the odd viscous part of Eq.~\eqref{eq:stress110909} due to no-penetration boundary conditions of the obstacle. However, there may still be a torque which enters the shear viscous part of Eq.~\eqref{eq:stress110909} as this term depends on the odd viscous part of the fluid profile. We find that up to second order in slip length, torque does not arise up to $\mathcal{O} (\hat k )$. Similarly, we find torque to vanish entirely in the previously considered cases of Stokes flow in a thin sheet and Stokes flow for a finite system size. 

\section{Discussion}
In this work we showed how slip can induce odd viscous flow past a cylinder for a variety of fluid systems. For Stokes flow, the corresponding drag and lift can be computed with the Lorentz reciprocal theorem up to first order in slip length, however lift force only arises at second order in slip length. We also went beyond the Stokes approximation by solving the Oseen equation up to second order in a harmonic expansion to find the flow and that the slip-dependence of lift force is qualitatively identical to that of the Stokes approximation. To the best of our knowledge, this is the first time that the role of inertia is considered for steady odd viscous flow past an obstacle.
\newline 
There is a myriad of possible directions for future study. Firstly, one could study Janus particles in the context of slip boundary conditions, i.e. to study  cylinders for which the slip length is not uniform around the cylinder. It is straightforward to generalize the results based on the Lorentz reciprocal theorem to the case where slip length is non-uniform. Similarly, one could consider cylinders which are not fully circular. Furthermore, there have been many recent works that study odd viscous flow around spheres \cite{PhysRevLett.127.048001,olvera,Khain_2022,reynolds2023dimensional,everts2023dissipative,khain2023trading,hosaka2023hydrodynamics}. In this case, odd viscous flow can exist even for no-slip boundary conditions as odd viscosity can no longer be absorbed into a modified pressure. All previous works on odd viscous flow past a sphere have been concerned with Stokes flow. It would be valuable to learn how inertia gives rise to an interplay with odd viscosity for flow past a sphere, with or without slip. Lastly, for bubbles, i.e. particles which do not have a constant volume, it is possible for odd viscosity to give rise to torque \cite{ganeshan2017odd}. One could explore the effects slip on the torque for such a bubble.
\acknowledgements
We thank Aleksander Głódkowski, Francesco Peña-Benitez, Piotr Surówka, Jonas Veenstra and especially Charlie Duclut for useful discussions.

\onecolumngrid

\appendix

\section{Oseen drag with slip}
\label{app:oseendragslip1}
In this appendix we consider the solution to the even Oseen equation in an expansion up to first harmonics to obtain the Oseen drag. We can then start from the ansatz \cite{tomotikaaa,lamb1932hydrodynamics,PhysRevB.99.035430}
    \begin{align} \label{eq:ansatz}
    \phi   &    = a U  \left( A_0  \log (r )  +    A_1 \frac{ a  }{ r} \cos( \theta ) \right)  ~~ ,   ~~ 
        \chi      =  U e^{  - k r \cos(\theta ) }      B_0 K_0 (k r ) ~~ , 
\end{align}
The constraints on the coefficients coming from the boundary conditions are given by
\begin{align} \label{eq:boundarycondition}
  \left[  M^{(0)}_{3 \times 3} + \hat{\lambda}  M^{(1)}_{3 \times 3}  \right]  \begin{bmatrix}
     A_0     &     B_0   &   A_1 \end{bmatrix}^T    =   \begin{bmatrix}
     0    & 
            1    &  
            -     1  
\end{bmatrix}^T ~~ , 
\end{align}
with
     \begin{align} 
 M^{(0)}_{3 \times 3}     &  = \begin{bmatrix}
      -1 & -\frac{1}{2 \hat{k}} & 0  \\
 0 & I_2 (\hat k) K_0 (\hat k)+ I_1 (\hat k)\left(\frac{K_0 (\hat k)}{\hat{k}}+K_1  (\hat k)\right) & 1  \\
 0 & -\frac{I_1 (\hat k)  K_0 (\hat k)}{\hat{k}} & 1 \end{bmatrix}
 ~~ ,  ~~  
M^{(1)}_{3 \times 3}     =   \begin{bmatrix} 0 & 0 & 0 \\
 0 & 0 & 0 \\
 0 & 2 I_2 (\hat k) K_0 (\hat k) & 4 
 \end{bmatrix}
 ~~   , 
\end{align}  
where $I_n(x)$ is the $n$th modified Bessel function of the first kind. Solving Eq.~\eqref{eq:boundarycondition} leads to the solutions
\begin{subequations}
    \begin{align}
     A_0   & = -\frac{2 \hat{\lambda}+1}{\hat{k} (4 \hat{\lambda}+1) I_1(\hat{k}) K_1(\hat{k})+ ( 2 \hat{\lambda}+1 )\hat{k} I_0(\hat{k}) K_0(\hat{k})}  ~~ ,  \\ 
          A_1   & =-\frac{(2 \hat{\lambda}+1) I_2(\hat{k}) K_0(\hat{k})+I_1(\hat{k}) K_1(\hat{k})}{(2 \hat{\lambda}+1) I_0(\hat{k}) K_0(\hat{k})+(4 \hat{\lambda}+1) I_1(\hat{k}) K_1(\hat{k})} ~~   , \\ 
          B_0 & = \frac{4 \hat{\lambda}+2}{(2 \hat{\lambda}+1) I_0(\hat{k}) K_0(\hat{k})+(4 \hat{\lambda}+1) I_1(\hat{k}) K_1(\hat{k})} ~~ .
\end{align}
\end{subequations}
The corresponding drag force is given by
\begin{align} \label{eq:unexpanded}
  C_D   =   \frac{2 \pi \eta_s (2 \hat \lambda +1) (2 I_1(\hat k) K_1(\hat k)+1)}{ (2 \hat \lambda +1) I_0(\hat k) K_0(\hat k)+  (4 \hat \lambda +1) I_1(\hat k) K_1(\hat k)} ~~ .
\end{align}
For low Reynolds number, this reduces to 
\begin{align}  \label{eq:lowreynoldsnumberdrag}
    C_D   =    \frac{4 \pi \eta_s}{ 1 -  \frac{1}{2 (2 \hat \lambda +1)}- \log (\frac{\hat k}{2})- \gamma_{EM}  }   + \mathcal{O} (\hat{k}) ~~ , 
\end{align}
where $\gamma_{EM}$ is the Euler-Mascheroni constant. Eq.~\eqref{eq:lowreynoldsnumberdrag} coincides with the result for drag force found in Ref.~\cite{LI2014211} using asymptotic matching. Lastly, it turns out that if one swaps the $A_1$-solution with the $B_1$-solution in the ansatz of Eq.~\eqref{eq:ansatz}, this modifies Eq.~\eqref{eq:unexpanded} but leaves Eq.~\eqref{eq:lowreynoldsnumberdrag} invariant.

\end{document}